\documentclass[twocolumn,prd,aps,floats,nofootinbib,showpacs]{revtex4}
\usepackage{epsfig}
\usepackage{graphicx}
\usepackage{color}
\newcommand{\gtrsim}{\,\rlap{\lower3.7pt\hbox{$\mathchar\sim$}}
\raise1pt\hbox{$>$}\,}
\newcommand{\lesssim}{\,\rlap{\lower3.7pt\hbox{$\mathchar\sim$}}
\raise1pt\hbox{$<$}\,}

\newcommand{\be}{\begin{equation}}
\newcommand{\ee}{\end{equation}}
\newcommand{\bea}{\begin{eqnarray}}
\newcommand{\eea}{\end{eqnarray}}
\def\d{{\rm d}}
\def\psim{\psi_{\rm M}}
\def\psimin{\psi_{\rm m}}
\def\df{d_{\rm f}}
\def\rhof{\rho_{\rm f}}
\def\alt{\raise0.3ex\hbox{$\;<$\kern-0.75em\raise-1.1ex\hbox{$\sim\;$}}}
\def\agt{\raise0.3ex\hbox{$\;>$\kern-0.75em\raise-1.1ex\hbox{$\sim\;$}}}
\def\sigv{\langle\sigma v\rangle}

\begin{document}

\title{Optimal angular window for observing Dark Matter annihilation
from the Galactic Center region: the case of gamma-ray lines.}
\author{Pasquale D.~Serpico}
\affiliation{Center for Particle Astrophysics, Fermi National
Accelerator Laboratory, Batavia, IL 60510-0500 USA}
\author{Gabrijela Zaharijas}
\affiliation{HEP Division, Argonne National Laboratory, 9700 Cass Ave., Argonne, IL 60439}

\begin{abstract}
Although the emission of radiation from dark matter annihilation is expected to
be maximized at the Galactic Center, geometric factors and the presence of point-like
and diffuse backgrounds make the choice of the angular window size to optimize the chance
of a signal detection a non-trivial problem.  We find that  the best strategy is to focus  on an annulus around the Galactic Center  
of $\sim 1^\circ$ to $\agt 30^\circ$, where the optimal size depends on the angular distribution of the signal and the backgrounds. 
Although our conclusions are general, we illustrate this point in the particular case of annihilation into two monochromatic photons in the phenomenologically most interesting range of energy $45\,{\rm GeV}\alt E\alt 80\,$GeV, which is of great
interest for the GLAST satellite. We find for example that Dark Matter models with sufficiently strong line annihilation signals, like the Inert
Doublet Model, may be detectable without or with reasonable boost factors.
\end{abstract}
\pacs{95.35.+d, 14.80.Cp, 98.70.Rz \hfill ANL-HEP-PR-08-16, FERMILAB-PUB-08-036-A}

\maketitle
\section{Introduction}\label{intro}
Although the existence of Dark Matter seems to be required by a wealth of astrophysical
and cosmological data (for reviews
see~\cite{Bertone:2004pz,Bergstrom:2000pn}), the identification of its physical nature remains still elusive. The most
popular candidates are the so-called weakly interacting massive particles (or WIMPs), 
believed
to be neutral and colorless thermal relics from the early universe with electroweak-scale masses ($m_X\sim 10-1000\,$GeV) and couplings. These generic assumptions seem indeed
to roughly reproduce the amount of DM needed in the Universe and in our Galaxy in particular, while arising at the same time naturally in many extensions of the Standard Model of Particle Physics.

Among the different methods proposed to identify WIMP dark matter (production in accelerators,
direct detection via recoil of nuclei underground, indirect detection via their annihilation or
decay products~\cite{Hooper:2008sn}), a particularly promising technique is the detection of
gamma rays from annihilation in the galactic halo. 
However, it is nowadays clear that the DM signal should be likely singled out of a larger 
astrophysical background, either point-like (the case of the Galactic Center) or diffuse.
In assessing the capability of detecting a given WIMP candidate it is important to
optimize the search strategy based on our current expectation for the energy range,
but also the angular distribution of the signal. This information proves actually to be at least as useful in guiding
DM searches as the energy spectrum and the choice of astrophysical targets.  Here we illustrate this
point with reference to the inner  Galaxy  region.
Although the emission of radiation per unit solid angle from dark matter annihilation is expected to
be maximized at the Galactic Center, geometric factors and the presence of point-like
and diffuse backgrounds make the choice of the window size to optimize the chance
of a signal detection a non-trivial problem.  To the best of our knowledge, this issue has 
been only raised within over-simplified assumptions for the signal and background~\cite{Stoehr:2003hf} or, by converse, treated in some technical studies by experimental groups devoted to dark matter searches (see for example \cite{poster}), with the background profile determined e.g. via a numerical model. In no case a comparative study for different profiles and background assumptions, which illustrates the physical origin of the optimal cuts to adopt, has been presented. This is the main task we address in this article.
 We shall prove that---in all the commonly considered
models for the DM distribution in our Galaxy and given the present knowledge of point-like and
diffuse backgrounds---the optimal strategy for DM searches is never to focus on the inner 
0.1$^\circ$-0.5$^\circ$ around the Galactic Center. Unless astrophysical processes dramatically enhance the DM
distribution in  the innermost region of our Galaxy, producing effectively a bright point-source, an annulus of radius from $\sim 1^\circ$ to $\agt 30^\circ$ is always to be preferred. The optimal size depends on the angular distribution of the signal and the backgrounds, and the improvement in the sensitivity reaches a factor of a few, thus affecting the prospects for the detectability  of specific models. To give an explicit example,
we shall refer mostly to the detection of gamma-ray lines from annihilation. Gamma-ray lines of energy 
$E=m_X$ would be an unambiguous signature of WIMPs  as DM particles populating our Galactic Halo,
but the two-photon channel is loop(or radiatively)-suppressed with respect to the main WIMP annihilation channels: fermion pairs, weak gauge bosons or higgses. Even models where the
line emission is enhanced, as the Inert Doublet Model (IDM)~\cite{Barbieri:2006dq,Gustafsson:2007pc}
are very challenging to detect via line emission. Yet, with an optimized choice of the angular size, the IDM model
may be detectable  without or with reasonable boost factors (i.e., enhancement of the flux with respect to the smooth halo signal due to the ``clumpiness" of DM). Similar considerations apply
for the more conventional continuum gamma-ray flux from DM annihilation, usually
considered a more promising channel.

This  paper is structured as follows. In Sec.~\ref{spectrum} we discuss the ingredients contributing
to the DM signal. In Sec.~\ref{backgrounds} we describe our model for the point-like and 
diffuse backgrounds, while Sec.~\ref{results} is devoted to present our results. In Sec.~\ref{concl}
we conclude.

\section{The Signal}\label{spectrum}

It is customary to write the number flux  of gamma rays (per energy and solid angle) produced in dark matter (DM) annihilations in
our Galactic halo as
\begin{equation}
\Phi^{\rm DM}(E,\psi) = \left[ \frac{\d N_{\gamma}}{\d E} \frac{\sigv}{8\pi m^2_X}\right]\,\df\,\rhof^2\,J(\psi)\,,\label{fluxDM}
\end{equation}
where the factor in square brackets depends only on the particle physics model: $\sigv$ is the WIMP annihilation cross section multiplied by the relative velocity of the two WIMPs 
(averaged over the WIMP velocity distribution), $m_X$ is the mass of the WIMP, 
$\d N_{\gamma}/\d E$ is the photon spectrum per annihilation.   If the dark matter is not its
own antiparticle (as assumed here), Eq.~(\ref{fluxDM})  should be
multiplied further by a factor 1/2 (if $X$ and $\bar{X}$ are equally abundant).

The factor outside the brackets only depends on the DM astrophysics, in particular on the distribution
of DM in the halo and on the position of the Solar System within the halo.
The quantities $\df$ and $\rhof$ are constants with dimensions of distance and
energy density respectively,  whose only purpose is of making $J$---which  encodes the angular dependence of
the DM signal---a dimensionless function.
Following~\cite{Bergstrom:1997fj},  we fix these parameters at $\df=8.5\,$kpc and $\rhof=0.3\,$GeV/cm$^3$ (indicative
 the galactocentric distance of the sun and of the local density of dark matter). The $J$-function is thus defined 
 in terms of the DM density $\rho$ as

\begin{equation}
J(\psi) \equiv\frac{1}{\df}\frac{1}{\rhof^2}  \int_{\rm{los}} \rho^2[r(\ell,\psi)] \d \ell,
\label{Jfunct}
\end{equation}
where
\begin{equation}
r(\ell,\psi)=\sqrt{r_\odot^2+\ell^2-2\,r_\odot\,\ell\cos\psi}\,. \label{rspsi}
\end{equation}
In the previous equations, $\psi$ is the angle between the direction in the sky and the
Galactic Center so that, in terms of galactic latitude $b$ and
longitude $l$,  $\cos\psi=\cos b\cos l$; $r_\odot\approx 8.0\,$kpc~\cite{Yao:2006px}
is the solar distance
from the Galactic Center; $\ell$ is the distance from the Sun along
the line-of-sight (l.o.s.). Note that we implicitly assumed that the halo is spherically
symmetric and that the angular dependence only comes from the off-center position
of the Sun in the halo, which implies for example the azimuthal symmetry of the signal 
(see e.g.~\cite{Hooper:2007be} and refs. therein for a discussion of sub-leading effects 
determining the angular distribution of the signal).
 A general class of spherically symmetric, smooth halo distributions can be fitted as
\begin{equation}
\rho(r)=\left(\frac{r_s}{r}\right)^\gamma
\frac{\rho_0}{[1+(r/r_s)^\alpha]^{(\beta-\gamma)/\alpha}},
\end{equation}
where $\rho_0$ is a normalization constant and $r_s$ is a characteristic  radius
below which the profile scales as $r^{-\gamma}$. Two of the most
well known profiles have been proposed by Navarro, Frenk and White
(NFW)~\cite{NFW} and Moore {\it et al.}~\cite{Moore}. Another fit, proposed by 
Kravtsov et al.~\cite{Kravtsov:1997dp}
mimics a flat core potential, which seems to be in better agreement with some observations.
More recent simulations (see e.g.~\cite{Power:2002sw,Navarro:2003ew,Reed:2003hp,Merritt:2005xc})
suggest that halo density profiles are better represented by a function with a
continuously-varying slope. At least above the resolved radius ($\sim 1\,$kpc) and for Galactic-size halos, the  newly proposed fitting formulae provide only marginal improvement with respect to the more
traditional ones, although it seems that the inner slope of a DM halo is very likely shallower that the Moore one, and probably shallower than the NFW one, too.
For illustrative purposes, we shall consider the Moore, NFW and Kravtsov profiles with the parameters  reported in Table I, since the former and the latter bracket the different cases, with the NFW (as
well as the paramterization used in~\cite{Merritt:2005xc}) falling in between. Similarly to Ref.~\cite{Yuksel:2007ac}, the normalizations are chosen so that the mass contained within
the solar circle provides the appropriate
DM contribution to the local rotational curves, such that 
$\rho_\odot\equiv\rho(r_\odot)$---the DM density at the solar distance
from the GC---yields to the canonical value  
$\rho_\odot= 0.3\,$GeV/cm$^3$ for the NFW
profile.
\begin{table}[!tbh]\label{TableII}
\caption{Halo parameters, see text and Ref.~\cite{Yuksel:2007ac}.}
\begin{tabular}{c|ccccc}
\hline\hline
Model    & $\alpha$ & $\beta$ & $\gamma$ &$\rho_\odot$ & $r_s$ \\
  & &  &  &[GeV$\,$cm$^{-3}$]& [kpc]\\
\hline\hline
Moore     & 1.5 & 3 & 1.5 &0.27 & 28\\
NFW          & 1.0 & 3 & 1.0 & 0.30 & 20\\
Kravtsov   & 2.0 & 3 & 0.4 & 0.47 & 10\\
\hline
\end{tabular}
\end{table}
Note that in none of the above-mentioned models astrophysical processes modifying the
pure DM profile are included. For example, ``adiabatic compression" may lead to the formation of a DM spike around the central supermassive black hole; mergers as well as  scattering on the dense stellar
cusp around the central black hole may destroy density enhancements, etc. While potentially relevant in determining the dark matter annihilation brightness of the Galactic Center, they only affect regions too close to the Galactic Center to be resolved angularly by present detectors. In general, they might
lead to a measurable gamma ray flux from the innermost angular bin even in presence of
relatively large astrophysical backgrounds, but in this case only the energy information is relevant
for the separation, for which the present discussion is irrelevant. We  not address these effects further; details can be found in the refs. cited in~\cite{Bertone:2004pz} or ref.~\cite{Fornasa:2007nr}.

The spectrum per annihilation depends on the nature of the WIMP, although the gamma-ray yield is 
often largely dominated by the result of the fragmentation and hadronization of the WIMP annihilation
products. Neutralinos in the MSSM, for example, typically annihilate to final states consisting of heavy 
fermions and gauge or Higgs bosons~\cite{jungman}. With the exception of the $\tau^+ \tau^-$ channel, each of these annihilation modes result in a very similar spectrum of gamma rays. 
In most of the following considerations, however, it is the angular distribution
of the DM radiation that determine the optimal angular window. To avoid
unnecessary complications, we shall illustrate our point considering mostly
the case of annihilation in two gammas, for which
\begin{equation}
\frac{\d N_{\gamma}}{\d E} =2\,b_{\gamma\gamma}\delta(E-m_X)\,,
\label{flux1}
\end{equation}
where $b_{\gamma\gamma}$ is the branching ratio for the 2-photon line. 
Although it would constitute a striking signature of a DM annihilation process,
in general the photon line is considered hard to detect, since it is loop-suppressed
with respect to the main annihilation modes~\cite{Bergstrom:1997fj}. This is not necessarily the
case in some DM models, as for example the Inert Doublet Model (IDM)~\cite{Barbieri:2006dq,Gustafsson:2007pc}.
In general, the line feature may be enhanced whenever: (i) the DM almost decouples
from fermions, while being mostly coupled to
gauge bosons; (ii) the DM candidate is in the mass
range $45\,{\rm GeV}\simeq m_Z/2<M_X< m_W\simeq{\rm 80}\,$GeV, the lower limit
coming from the bound on the $Z^0$-width measurement and the upper limit to avoid
the annihilation mode $X\,X\to W^{+}W^{-}$ by kinematical constraints (other cases
where a prominent line emission is possible at lower or higher energies have been
treated respectively in~\cite{Pullen:2006sy} and~\cite{Bringmann:2007nk}).
Interestingly, this is probably the best region where the GLAST satellite~\cite{GLASTurl}
has
a reasonable chance to detect a line emission and at the same time, potentially
interesting for the next generation of low-threshold atmospheric cherenkov telescopes (ACTs),
like CTA~\cite{CTA} or AGIS~\cite{AGIS}.   In the following, for GLAST we shall use 
an integrated area exposure over time of $10^4\,{\rm cm}^2\,{\rm yr}$, corresponding to about
5 years of observations, and an energy resolution of $\pm 7\%$ at $1\sigma$. For ACTs, we shall assume an effective array area of $A_{\rm eff}=1\,$km$^2$ , energy resolution of $15\%$ and 200 hour observations, with a $2^\circ\times 2^\circ$ field of view.

Whenever needed, we shall compare our results with two benchmark models
for IDM candidates among the ones considered in \cite{Gustafsson:2007pc}, see Table~\ref{TableII}.
Also, to get a feeling of how things change if the continuum gamma-ray spectrum is considered instead,
we shall also report the results obtained when a fiducial case of a 100 GeV MSSM neutralino mainly
annihilating into $W^{+}W^{-}$ is considered.

\begin{table}[!tbh]\label{TableII}
\caption{Benchmark models in the IDM scenario, see~\cite{Gustafsson:2007pc}.}
\begin{tabular}{c|ccc}
\hline\hline
 Model       & $m_X$ & $\sigv$ & $b_{\gamma\gamma}$ \\
       & [GeV] & [cm$^3$/s] & \\\hline\hline
$I$       & $70$ & $1.6\times 10^{-28}$ & $0.36$ \\
\hline
$II$       & $50$ & $8.2\times 10^{-29}$ & $0.29$ \\
\hline
\end{tabular}
\end{table}
%

\section{The Backgrounds} \label{backgrounds}

The Galactic Center is a complex region of the sky at all wavelengths, the gamma-ray window being no exception. In this section, we discuss how, in our analysis, we treat the backgrounds for dark matter searches due to known and unknown astrophysical sources of gamma rays. Since we focus on an extended signal, diffuse (or unresolved) backgrounds
are our main concern. However, to illustrate how effective is a large field-of-view search, we shall compare it with
the sensitivity achieved when focusing on the inner region around the Galactic Center, when at least one
astrophysical point-like background is known to exist. This is the relatively bright, very high-energy gamma ray source observed 
by HESS, MAGIC, WHIPPLE, and CANGAROO-II~\cite{GCtev}. This source is consistent with point-like emission and is located at $l = 359^\circ 56^\prime 41.1^{\prime\prime}\pm 6.4^{\prime\prime}$ (stat.), 
$b = -0^\circ2^{\prime}39.2^{\prime\prime}\pm 5.9^{\prime\prime}$ (stat.) with a systematic pointing error of
28$^{\prime\prime}$ \cite{van Eldik:2007yi}. It appears to be coincident with the position of Sgr A$^{\star}$, the black hole constituting 
the dynamical center of the Milky Way. Following the measurements of HESS, we describe the spectrum of this source as a power-law given by:
\begin{equation}
\Phi^{\rm GC} = 1.0 \times 10^{-8} \left({ {E}\over{{\rm GeV}}}\right)^{-2.25} {\rm GeV}^{-1} \, {\rm cm}^{-2} \, {\rm s}^{-1}.
\end{equation}
At energies below $\sim 200$ GeV, the spectrum of this source has not yet been measured. GLAST, however, will be capable of measuring the spectrum of this source at energies below the  thresholds of HESS and other ACTs. 

In addition to the HESS source, a yet unidentified source has been detected by EGRET in the GeV range, approximately 0.2$^{\circ}$ away from the dynamical center of our galaxy~\cite{dingus,pohl}. It has not been detected so-far
at the energies explored by HESS and other ACTs, which means that its power-law spectrum cuts-off between $\sim$10 and $\sim$100 GeV.   Although it should be included in more realistic analyses of the DM signal from the inner galactic regions~\cite{Dodelson:2007gd}, in the following we neglect it, thus overestimating the diagnostic power of a narrow field
of view search and providing a {\it conservative} estimate of the improvement achieved by a large angle search.

The overall diffuse gamma-ray radiation can be qualitatively divided
into a galactic and an extragalactic contribution. Since the latter
is not simply the isotropic part of the flux,  the separation of
these two components can be done at present only assuming a specific
model for the production of secondaries by cosmic rays in the
galactic disk and halo. (However, a measurement of the cosmological
Compton-Getting effect that should be achievable for GLAST would
provide a model-independent way to separate the two
contributions~\cite{Kachelriess:2006aq}). A significant fraction of
the quasi-isotropic component, especially in the GeV range, may be
due to high-latitude galactic emission coming from processes in the
magnetized halo of the Milky Way. We employ here a fit of the galactic
diffuse flux proposed in~\cite{Bergstrom:1997fj} and calibrated on
EGRET data around the GeV~\cite{Hunter:1997},
\begin{widetext}
\be \label{spectrumGal} I_{\rm gal}(E)= N_0(l,b) \times 10^{-6}
\left(\frac{E}{{\rm GeV}}\right) ^{-\alpha}\,{\rm cm}^{-2}{\rm
s}^{-1}{\rm sr}^{-1}{\rm GeV}^{-1} \,, \ee
where the arguments $l,b$ are in degrees, $-180^\circ\leq l\leq 180^\circ$
and $-90^\circ\leq b\leq 90^\circ$, $\alpha\simeq 2.7$, and 
\begin{equation}   \label{linsys1}
N_{0}(l,b)=\left\{
\begin{array}{cc}
\frac{85.5}{\sqrt{1+(l/35)^2}\sqrt{1+[b/(1.1+0.022\,|l|)]^2}}+0.5 \,,
&\:\:|l|\geq 30^{\circ}\\
\frac{85.5}{\sqrt{1+(l/35)^2}\sqrt{1+(b/1.8)^2}}+0.5 \,,
&\:\:|l|\leq 30^{\circ}
\end{array}
\right..
\end{equation}

The EGRET collaboration derived the intensity of the extragalactic
gamma-ray flux as~\cite{Sreekumar:1997un}
\begin{equation}
I_{\rm ex}(E)=k_0\times 10^{-6}\left(\frac{E}{0.451 {\rm
GeV}}\right) ^{-\beta} {\rm cm}^{-2}{\rm s}^{-1}{\rm
sr}^{-1}{\rm GeV}^{-1} \, , \label{spectrum98}
\end{equation}
with $\beta=2.10\pm0.03$, $k_0=(7.32\pm 0.34)$ and the fits is valid from $E\sim\,$10~MeV to $E\sim\,$100~GeV.
\end{widetext}

In the following, we shall consider two models of diffuse backgrounds: a ``conservative'' model assuming 
$\alpha=2.7$ in Eq.~(\ref{spectrumGal}) and $k_0=7.32$ in Eq.~(\ref{spectrum98}), and an ``optimistic'' model,
where a harder spectrum $\alpha=3.0$ is assumed for the Galactic background and $k_0=3.66$. Based
on presently available models, one expects indeed that GLAST should resolve roughly half of the extragalactic 
background~\cite{Stecker:2001dk}; analogously, one expects that unresolved sources
contribute significantly to the diffuse Galactic spectrum as measured by EGRET, and once detected and removed
by GLAST the Galactic background should be steeper (see e.g. discussion in~\cite{Pullen:2006sy}).

Finally, for ACTs the dominant background is the isotropic cosmic ray one. At energies well below $\sim 100\,$GeV
one expects the cosmic-ray electron background to play a dominant role. In the following, we account for these
backgrounds following the formulae reported in~\cite{Bergstrom:1997fj}.

\section{Results}\label{results}
First, let us repeat a simple argument already sketched in~\cite{Stoehr:2003hf}.
The DM signal coming from integrating over a solid angle $\Delta\Omega$ is 
proportional to $\langle J(\Omega)\rangle\Delta\Omega$, where 
\begin{equation}
\langle J(\Omega)\rangle\equiv \frac{1}{\Delta\Omega}\int_{\Delta\Omega}J\,\d \Omega\,.
\end{equation}
For an annulus around the GC with inner opening semi-angle $\psimin$ and outer one $\psim$, the previous formula
reduces to
\begin{equation}
\langle J\rangle= \frac{1}{\cos\psimin-\cos\psim}\int_{\psimin}^{\psim}J(\psi)\sin\psi\d \psi\,.\label{Javer}
\end{equation}

The significance $S/N$  of a Signal $S$ in presence of a background $B$ is defined as
\begin{equation}
\frac{S}{N}\equiv \frac{S}{\sqrt{S+B}} \label{significance}
\end{equation}
The dark matter signal goes like $S\propto \langle J\rangle\,\Delta\Omega$, while for an isotropic 
background $B\propto \Delta\Omega$. In the limit $S\ll B$ (which is likely to hold in all realistic DM searches), we have
\begin{equation}
\frac{S}{N}\propto  \langle J\rangle\sqrt{\Delta\Omega}\sim \langle J\rangle\psim\,. \label{significance2}
\end{equation}

\begin{figure}[t]
\vspace*{-5mm}
\centering
\epsfig{figure=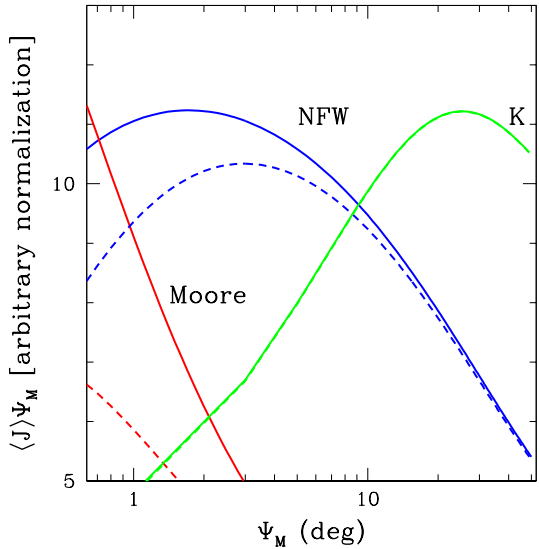,width =1.\columnwidth,angle=0}
 \caption{\label{Fig1} 
Plot of  the function $\langle J\rangle\psim$ defined in Eq.~(\ref{Javer}) (proportional to the significance of a DM signal for an isotropic background in the limit $S\ll B$) for the three models considered in Table I: the solid lines assume
$\psimin=0.1^\circ$, the dashed lines  $\psim=0.3^\circ$; the leftmost (red) lines are for the Moore profile, the rightmost (green)
lines for the Kravtsov, in the middle (blue) lines for the NFW profile.}
\end{figure}

In Fig.~\ref{Fig1}, we plot this function vs. $\psim$ for the three models considered in Table II, assuming
$\psimin=0.1^\circ$ (solid lines) or $\psimin=0.3^\circ$ (dashed lines). Note the exact choice of the inner
bound is irrelevant for cored profiles: the peak of this function is above 20$^\circ$. Even for profiles as 
cusped as the NFW (or more) the peak of this function is at a few degrees. Of course, the Moore profile 
gives a formally divergent signal when $\psimin\to 0$, so the inner angle cutoff does matter.

\begin{figure}[!htb]
\centering
\epsfig{figure=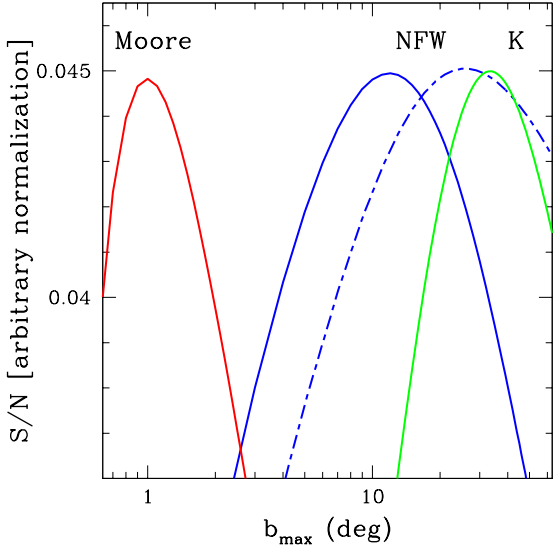,width =1.\columnwidth,angle=0}
 \caption{\label{Fig2} 
{\it relative}  Signal/Noise vs. annulus size $b_{\rm max}$ (we equate their maximum value to ease the comparison) for a region $0.4^\circ<|b|<b_{\rm max}$, $0^\circ<|l|<l_{\rm max}=b_{\rm max}$ for the
continuum spectrum of a 100 GeV MSSM neutralino annihilating into $W^{+}W^{-}$ (dot-dashed line,
NFW profile) and for the fiducial IDM model I of Tab. II for the Moore, NFW and Kravtsov profile
(solid lines respectively from the left to the right).}
\end{figure}

Figure~\ref{Fig1} already shows that the $S/N$ is typically maximized for values of $\psim \gg \psimin$. However, the diffuse backgrounds are not $\{l,b\}$-independent, but really decline away from the Galactic Plane. When using the models for the galactic and extragalactic background introduced in the previous section, we see how this effect is appreciable. In Fig.~\ref{Fig2} , in order to make the comparison among the different models easier, we plot the rescaled  $S/N$ vs.  $b_{\rm max}$ (in
such a way that their maximum values are equal)
for a region $0.4^\circ<|b|<b_{\rm max}$, $0^\circ<|l|<l_{\rm max}=b_{\rm max}$ for the fiducial IDM model I of Tab. II, and for the
continuum spectrum of a 100 GeV MSSM neutralino annihilating into $W^{+}W^{-}$. In the former case, we
consider the 1-$\sigma$ smoothing of the gamma-line due to the GLAST energy resolution, while for the continuum we
actually show the S/N in the most significant energy bin (which happens to fall in the  $\sim 1\,$GeV range).  
The inner cut has been chosen to $0.4^\circ$ to avoid contamination from known point sources close to the GC. Note how the peak of $S/N$ moves to larger angles in all cases compared with Fig.~\ref{Fig1}, developing a maximum at finite angle also for the Moore profile.
For the continuum spectrum the peak of the significance moves to larger angles, due to the fact that the isotropic component of the background is harder, and thus at the lower energies relevant for this signal the overall background is dominated by the Galactic background which falls   rapidly with $b$.  It is also worth nothing that even for  heavy dark matter particles (TeV mass
scale)  the continuum photon spectrum peaks at or below about  30 GeV. Since the $S/N$ for continuum spectra is dominated by its value at the peak, the angular window results reported in Figs. 1 and 2 are really more general than the case they were discussed for, reflecting the situation for the continuum spectrum from DM in all the interesting mass range. 

\begin{figure}[!htb]
\centering
\epsfig{figure=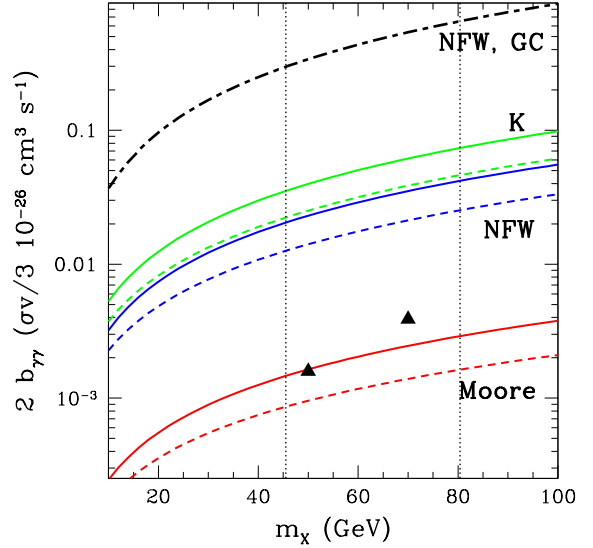,width =1.\columnwidth,angle=0}
 \caption{\label{Fig3} Plot of the curves with $S/N=3$ in the $10-100$ GeV range, with vertical
 lines in correspondence of $m_Z/2$ and $m_W$. The dot-dashed line shows the sensitivity achieved if one only includes the inner $0.1^\circ$  region around the GC  for the NFW profile.
The couples of solid/dashed lines (solid for conservative background, dashed for optimistic one) represent instead the sensitivity for the optimal annulus size for each model: from top to bottom, for
Kravtsov, NFW and Moore profiles, respectively.}
\end{figure}

In Fig.~\ref{Fig3} we show the curves with $S/N=3$ for the branching ratio into two gammas, normalized in terms
of the fiducial annihilation  cross section value $\sigv=3\times 10^{-26}\,$cm$^3$/s in the phenomenologically
most interesting mass range below $m_W$. The triangles represent the fiducial points of the IDM model
reported in Table II, and for the window size that maximizes $S/N$ (respectively $b_{\rm max}\sim1.5^\circ$, 15$^\circ$ and 30$^\circ$, for Moore, NFW and Kravtsov profiles of Tab. I). 
The solid lines are for the conservative background estimate, the dashed lines for the optimistic one.
It is interesting to note that for the Moore profile, the fiducial models should be detectable at the $3\,\sigma$
level without need for boost factors. For the NFW case,  boost factors of order 10 should suffice.
Even for the unfortunate case of a Kravtsov profile, a boost factor $\sim20$ should be sufficient.
These values appear to be viable in cold DM cosmologies, see e.g.~\cite{Strigari:2006rd}.
Thus, our conclusions are more promising that what envisioned in~\cite{Gustafsson:2007pc}, where
boost factors $\agt 100$ were considered.
Also, the top dot-dashed line shows the sensitivity achieved if one only includes the inner $0.1^\circ$ 
region around the GC, where the background is dominated by the central source. These limits,
comparable with the results reported in \cite{gabi}, show an evident  improvement. 

It is also worth commenting on the fact that, according to our estimate, although ACTs are not
optimized to search for {\it line} signals in the range below 100 GeV, the performance of the
next generation, km$^2$ instruments presently being considered may be comparable or better than GLAST. For example, in accordance with existing estimates~\cite{ACTwp}, for the fiducial
exposure parameters we have used and the NFW profile we find a sensitivity a factor $\sim$1.5 better in the parameter space of Fig. 3, despite their worse energy resolution. It is also fortunate that the ground-based instruments, that have
necessarily a much smaller field of view with respect to satellite observatories like GLAST,
do not suffer too much from limited field of view, at least for relatively cusped profiles. The dominant backgrounds of ACTs are indeed isotropically distributed,  and for this case the optimal annulus size is $\lesssim 2^\circ\div 3^\circ$ (for cusped profiles), within the field of view range considered by present 
designs for future ACT.

\section{Discussion and Conclusions}\label{concl}
A large field of view instrument like GLAST promises to revolutionize the field
of high energy gamma ray astrophysics. But it could also turn into a discovery
instrument for DM, if it is made by WIMPs which annihilate into standard model
particles in the halo of our galaxy. Since the DM signal grows like the square of its
density, it is usually assumed that the Galactic Center, where the DM density is
expected to be the highest, is the most promising region to focus on.
However, two arguments conspire to modify this
naive expectation: i) the presence of astrophysical point-like backgrounds close to the GC;
ii) geometric considerations, in particular related to the different angular shape
of the DM signal and astrophysical backgrounds.

It is thus important to  establish the level of the diffuse galactic and extragalactic diffuse 
backgrounds, a task which will be accomplished with exquisite precision by GLAST itself.
Yet, by employing simple models based on EGRET data, we showed that it is {\it always} to be preferred a window size $>1^\circ$ around the GC, even for profiles as steep as the Moore one.
The flatter the profile, the larger the optimal window size: for example we find an optimal
galactic latitude cut $b_{\rm max}\agt 10^\circ$ for NFW, $b_{\rm max}\agt 30^\circ$ for Kravtsov profile.  The discovery of additional point-like sources is not expected to change things dramatically, since
especially at high energies the good angular resolution of GLAST will allow one to remove the quasi-totality of the resolved point-source contaminations with only a minor loss in the solid angle of the diffuse signal. 

When these geometric effects are taken into account and the search strategy is adjusted consistently, the prospects for DM detection improve considerably. 
For example, we found that models with sufficiently strong line annihilation signals, like the Inert
Doublet Model, may be detectable without or with still reasonable boost factors, of the order
of 10 even for Kravtsov-like profiles. Also, the perspectives to detect a DM signal become less
sensitive to the unknown profile of the DM halo. In particular the $S/N$ of quasi-cored profiles is only a factor $\sim 2$ worse than for the NFW case. 
We also showed with one example that our considerations remain true 
also for the widely considered case of continuum  gamma-ray spectrum from DM annihilation. Given the generality of these considerations, we believe they should be taken
into account when estimating the reach of DM searches via gamma-ray telescopes.\\
{}\\
{\bf Acknowledgments.}
Authors are grateful to James Buckley and Dan Hooper for careful reading of the manuscript and useful comments. This work is supported in part by the US Department of Energy, Division of High Energy Physics, under Contract DE-AC02-06CH11357. P.S. is supported by the US Department of Energy and by NASA grant NAG5-10842. Fermilab is operated by Fermi Research Alliance, LLC under Contract No.~DE-AC02-07CH11359 with the United States Department of Energy. 

\begin{thebibliography}{100}

\bibitem{Bertone:2004pz}
  G.~Bertone, D.~Hooper and J.~Silk,
  Phys.\ Rept.\  {\bf 405}, 279 (2005)
  [hep-ph/0404175].


\bibitem{Bergstrom:2000pn}
  L.~Bergstrom,
  Rept.\ Prog.\ Phys.\  {\bf 63}, 793 (2000)
  [hep-ph/0002126].

\bibitem{Hooper:2008sn}
  D.~Hooper and E.~A.~Baltz,
  arXiv:0802.0702 [hep-ph].


\bibitem{Stoehr:2003hf}
  F.~Stoehr, S.~D.~M.~White, V.~Springel, G.~Tormen and N.~Yoshida,
  Mon.\ Not.\ Roy.\ Astron.\ Soc.\  {\bf 345}, 1313 (2003)
  [astro-ph/0307026].
  
 \bibitem{poster} \url{http://confluence.slac.stanford.edu/download/attachments/19303/HEAD_Poster.pdf?version=1}
 
 \bibitem{Barbieri:2006dq}
  R.~Barbieri, L.~J.~Hall and V.~S.~Rychkov,
  Phys.\ Rev.\  D {\bf 74}, 015007 (2006)
  [hep-ph/0603188].


\bibitem{Gustafsson:2007pc}
  M.~Gustafsson, E.~Lundstrom, L.~Bergstrom and J.~Edsjo,
  Phys.\ Rev.\ Lett.\  {\bf 99}, 041301 (2007)
  [astro-ph/0703512].


\bibitem{Bergstrom:1997fj}
  L.~Bergstrom, P.~Ullio and J.~H.~Buckley,
  Astropart.\ Phys.\  {\bf 9}, 137 (1998)
  [astro-ph/9712318].


\bibitem{Yao:2006px}
  W.~M.~Yao {\it et al.}  [Particle Data Group],
  J.\ Phys.\ G {\bf 33}, 1 (2006).

\bibitem{Hooper:2007be}
  D.~Hooper and P.~D.~Serpico,
  JCAP {\bf 0706}, 013 (2007)
  [astro-ph/0702328].


\bibitem{NFW} J. F. Navarro {\it et al.},
Astrophys. J. 462 563 (1996).

\bibitem{Moore} B. Moore {\it et al.},
Phys.\ Rev.\ D {\bf64} 063508 (2001).

\bibitem{Kravtsov:1997dp}
  A.~V.~Kravtsov, A.~A.~Klypin, J.~S.~Bullock and J.~R.~Primack,
  Astrophys.\ J.\  {\bf 502}, 48 (1998)
  [astro-ph/9708176].

\bibitem{Power:2002sw}
  C.~Power {\it et al.},
  Mon.\ Not.\ Roy.\ Astron.\ Soc.\  {\bf 338}, 14 (2003)
  [astro-ph/0201544].

\bibitem{Reed:2003hp}
  D.~Reed {\it et al.},
  Mon.\ Not.\ Roy.\ Astron.\ Soc.\  {\bf 357}, 82 (2005)
  [astro-ph/0312544].

\bibitem{Navarro:2003ew}
  J.~F.~Navarro {\it et al.},
  Mon.\ Not.\ Roy.\ Astron.\ Soc.\  {\bf 349}, 1039 (2004)
  [astro-ph/0311231].

\bibitem{Merritt:2005xc}
  D.~Merritt, J.~F.~Navarro, A.~Ludlow and A.~Jenkins,
  Astrophys.\ J.\  {\bf 624}, L85 (2005)
  [astro-ph/0502515].

\bibitem{Yuksel:2007ac}
  H.~Yuksel, S.~Horiuchi, J.~F.~Beacom and S.~Ando,
  Phys.\ Rev.\  D {\bf 76}, 123506 (2007)
  [arXiv:0707.0196 [astro-ph]].

\bibitem{Fornasa:2007nr}
  M.~Fornasa and G.~Bertone,
  arXiv:0711.3148 [astro-ph].

\bibitem{jungman}
  G.~Jungman, M.~Kamionkowski and K.~Griest,
  Phys.\ Rept.\  {\bf 267}, 195 (1996).

\bibitem{Pullen:2006sy}
  A.~R.~Pullen, R.~R.~Chary and M.~Kamionkowski,
  Phys.\ Rev.\  D {\bf 76}, 063006 (2007)
  [astro-ph/0610295].


\bibitem{Bringmann:2007nk}
  T.~Bringmann, L.~Bergstrom and J.~Edsjo,
  arXiv:0710.3169 [hep-ph].

\bibitem{GLASTurl}
\url{http://www-glast.slac.stanford.edu/}

\bibitem{CTA}
\url{http://www.mpi-hd.mpg.de/hfm/CTA/}

\bibitem{AGIS}
\url{http://cherenkov.physics.iastate.edu/wp/}

\bibitem{GCtev}
F.~Aharonian {\it et al.}  [The HESS Collaboration],
 astro-ph/0408145;
  J.~Albert {\it et al.}  [MAGIC Collaboration],
  Astrophys.\ J.\  {\bf 638}, L101 (2006)
  [astro-ph/0512469];
  K.~Kosack {\it et al.}  [The VERITAS Collaboration],
  Astrophys.\ J.\  {\bf 608}, L97 (2004)
  [astro-ph/0403422];
 K.~Tsuchiya {\it et al.}  [CANGAROO-II Collaboration],
  Astrophys.\ J.\  {\bf 606}, L115 (2004)
  [astro-ph/0403592].
  
\bibitem{van Eldik:2007yi}
  C.~van Eldik, O.~Bolz, I.~Braun, G.~Hermann, J.~Hinton and W.~Hofmann  [for
                  the HESS Collaboration],
  arXiv:0709.3729 [astro-ph].
  
  
\bibitem{dingus}
  D.~Hooper and B.~L.~Dingus,
  Phys.\ Rev.\ D {\bf 70}, 113007 (2004)
  [astro-ph/0210617];
  D.~Hooper and B.~Dingus,
Proc.~of the 34th COSPAR Scientific Assembly, Houston, Texas (2002),  
  astro-ph/0212509.

\bibitem{pohl}
M.~Pohl,
Astrophys.\ J.\  {\bf 626}, 174 (2005).

\bibitem{Dodelson:2007gd}
  S.~Dodelson, D.~Hooper and P.~D.~Serpico,
  Phys.\ Rev.\  D {\bf 77}, 063512 (2008)
  [arXiv:0711.4621 [astro-ph]].
  

\bibitem{Kachelriess:2006aq}
  M.~Kachelrie{\ss} and P.~D.~Serpico,
  Phys.\ Lett.\  B {\bf 640}, 225 (2006)
  [astro-ph/0605462].

  \bibitem{Hunter:1997}
S.~D.~Hunter {\it et al.}  [EGRET Collaboration],
Astrophys.\ J.\  {\bf 481}, 205 (1997).


\bibitem{Sreekumar:1997un}
P.~Sreekumar {\it et al.}  [EGRET Collaboration],
Astrophys.\ J.\  {\bf 494}, 523 (1998)  [astro-ph/9709257].



\bibitem{Stecker:2001dk}
  F.~W.~Stecker and M.~H.~Salamon,
  astro-ph/0104368.
  
  
\bibitem{Strigari:2006rd}
  L.~E.~Strigari, S.~M.~Koushiappas, J.~S.~Bullock and M.~Kaplinghat,
  Phys.\ Rev.\  D {\bf 75}, 083526 (2007)
  [astro-ph/0611925].

\bibitem{gabi}
  G.~Zaharijas and D.~Hooper,
  Phys.\ Rev.\  D {\bf 73}, 103501 (2006).
  [astro-ph/0603540].
  
  \bibitem{ACTwp} J. H. Buckley {\it et al.}, 
whitepaper for ``Dark Matter Searches with a Future VHE -Ray Observatory'', in preparation.


    \end{thebibliography}
  \end{document}